\PassOptionsToPackage{maxnames=4}{biblatex}

\documentclass[biblatex]{lni}
\usepackage{booktabs}
\usepackage{todonotes}

\usepackage[]{blindtext}

\usepackage{tcolorbox}
\usepackage{tikz}
\newcommand*\circled[1]{\tikz[baseline=(char.base)]{
            \node[shape=circle,draw,inner sep=1pt] (char) {#1};}}
\usepackage{svg}
\graphicspath{{images/}}

\begin{document}
\selectlanguage{english}

\title[Towards View-based Development of Quantum Software]{Towards View-based Development of Quantum Software}
\author[1]{Joshua Ammermann}{joshua.ammermann@kit.edu}{0000-0001-5533-7274}
\author[2]{Wolfgang Mauerer}{wolfgang.mauerer@othr.de}{0000-0002-9765-8313}
\author[1]{Ina Schaefer}{ina.schaefer@kit.edu}{0000-0002-7153-761X}
\affil[1]{KIT\\ Institute of Information Security and Dependability (KASTEL)\\Karlsruhe\\ Germany}
\affil[2]{Technical University of Applied Sciences Regensburg and Siemens Technology\\ Regensburg/Munich\\Germany}
\maketitle

\begin{abstract}
Quantum computing is an interdisciplinary field that relies on the expertise of many different stakeholders. 
The views of various stakeholders on the subject of quantum computing may differ, thereby complicating communication.
To address this, we propose a view-based quantum development approach based on a Single Underlying Model (SUM)
and a supporting quantum Integrated Development Environment (IDE).
We highlight emerging challenges for future research.
\end{abstract}

\section{Introduction}

Quantum computing is an interdisciplinary field that draws upon expertise ranging from the concrete quantum mechanical effects to the systematic development of quantum software.
Many different stakeholders, each with a distinct background, are involved, such as mathematicians, physicists, engineers, and computer scientists.
The mental models of stakeholders and their preferred view on quantum computing and programming may differ, which presents a significant challenge in terms of interdisciplinary collaboration.
Stakeholders want to work with their preferred view of a quantum program while being supported by a model- and view-based approach that combines mutual views.

There are some established concepts and tools that support one or more views of quantum computing.
Such views include mathematical formulas (linear algebra), quantum circuits, and Quantum Programming Languages (QPLs).
Many QPLs are embedded in a classical host language (e.g., Qiskit is embedded in Python), so they can reuse its tooling. 
For other languages such as Q\#~\cite{svoreEnablingScalableQuantum2018} a Visual Studio Code plugin is available as part of the Azure Quantum Development Kit\footnote{\url{https://github.com/microsoft/qsharp}}.
While this tooling provides comfortable editors (e.g., with syntax highlighting), it lacks support for adequate quantum views other than the QPL.
Q-UML~\cite{perez-delgadoQuantumSoftwareModeling2022} extends the Unified Modeling Language (UML) with quantum elements, such as quantum classes or variables, but fails to address quantum-specific characteristics adequately~\cite{aliNeedQuantumOrientedParadigm2023}.
First Integrated Development Environments (IDEs) for quantum software are under development.
Quirk\footnote{\url{https://github.com/Strilanc/Quirk}} and IBM Composer\footnote{\url{https://quantum.ibm.com/composer}} both provide a circuit editor, while IBM Composer has an additional OpenQASM / Qiskit view, but only supports one-way-editing for Qiskit code.
The QuantumPath IDE~\cite{heviaQuantumPathQuantumSoftware2022} provides disjoint annealing, circuit, and code editors.
The Classiq IDE~\cite{minerbiQuantumSoftwareDevelopment2022} provides a high-level description in a quantum modeling language that can be synthesized into circuits, but this is merely a unidirectional transformation.
Thus, existing IDEs still mostly support only one-way-editing / unidirectional transformations between views.
These are not sufficient to support a collaborative workflow between an interdisciplinary group of stakeholders.

In classical software engineering, a model is an abstraction of a real system towards specific aspects of interest.
Model-based techniques use models as the central artifact, facilitating the creation of domain-specific views and languages~\cite{wasowskiDomainSpecificLanguagesEffective2023}.
View-based development techniques allow the organization and generation of different views through the use of a (Virtual) Single Underlying Model (V-SUM)~\cite{meierClassifyingApproachesConstructing2020}.
A V-SUM ensures consistency of underlying model(s) and derived views, and enables collaborative editing.
These techniques are being adopted in interdisciplinary fields, such as cyber-physical systems~\cite{reussnerConsistencyViewBasedDevelopment2023}, and appear suitable for addressing the needs of the interdisciplinary field of quantum computing.

A comprehensive approach to view-based development of quantum software is yet to be established.
Such an approach enables stakeholders to view and edit different views, which will be held consistent automatically.
Thus, the research challenge is:
\textbf{How can a view-based approach to quantum software development be defined and realized?}




\section{View-based Quantum Development Approach}

\begin{figure}[b]
    \centering
    \includegraphics[width=0.75\textwidth]{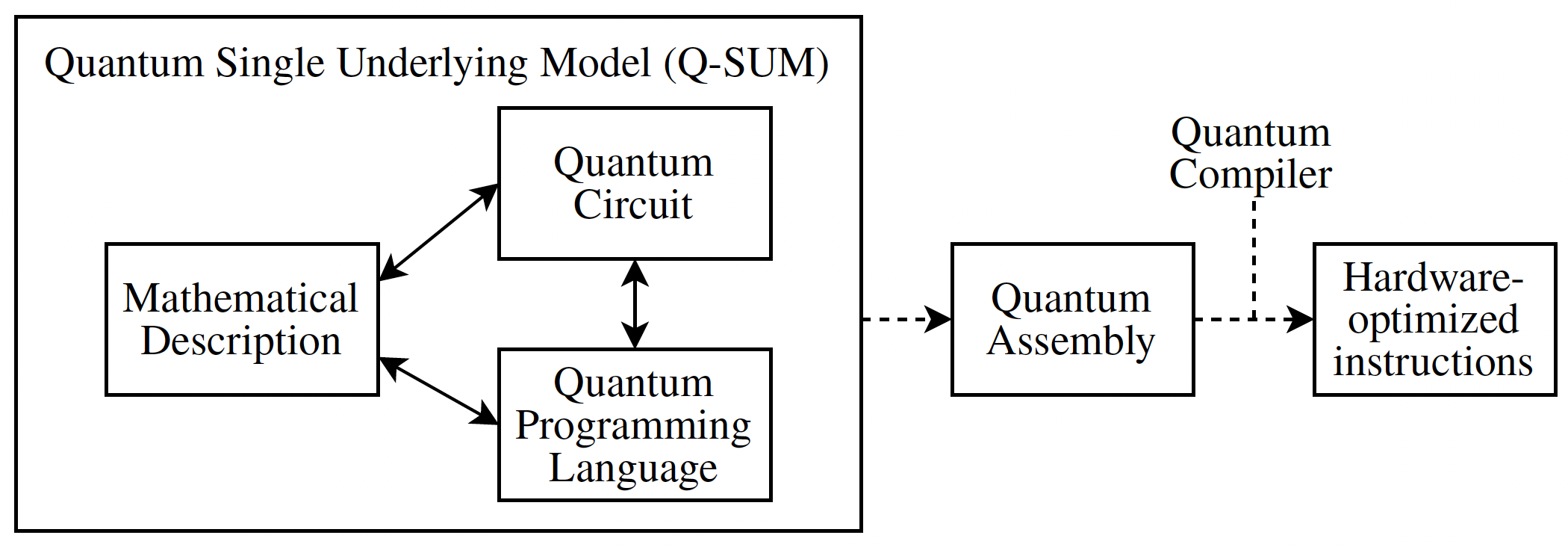}
    \caption{View-based Quantum Development approach relying on a Q-SUM}
    \label{fig:concept}
\end{figure}

\begin{figure}[b]
    \centering
    \includegraphics[width=\textwidth]{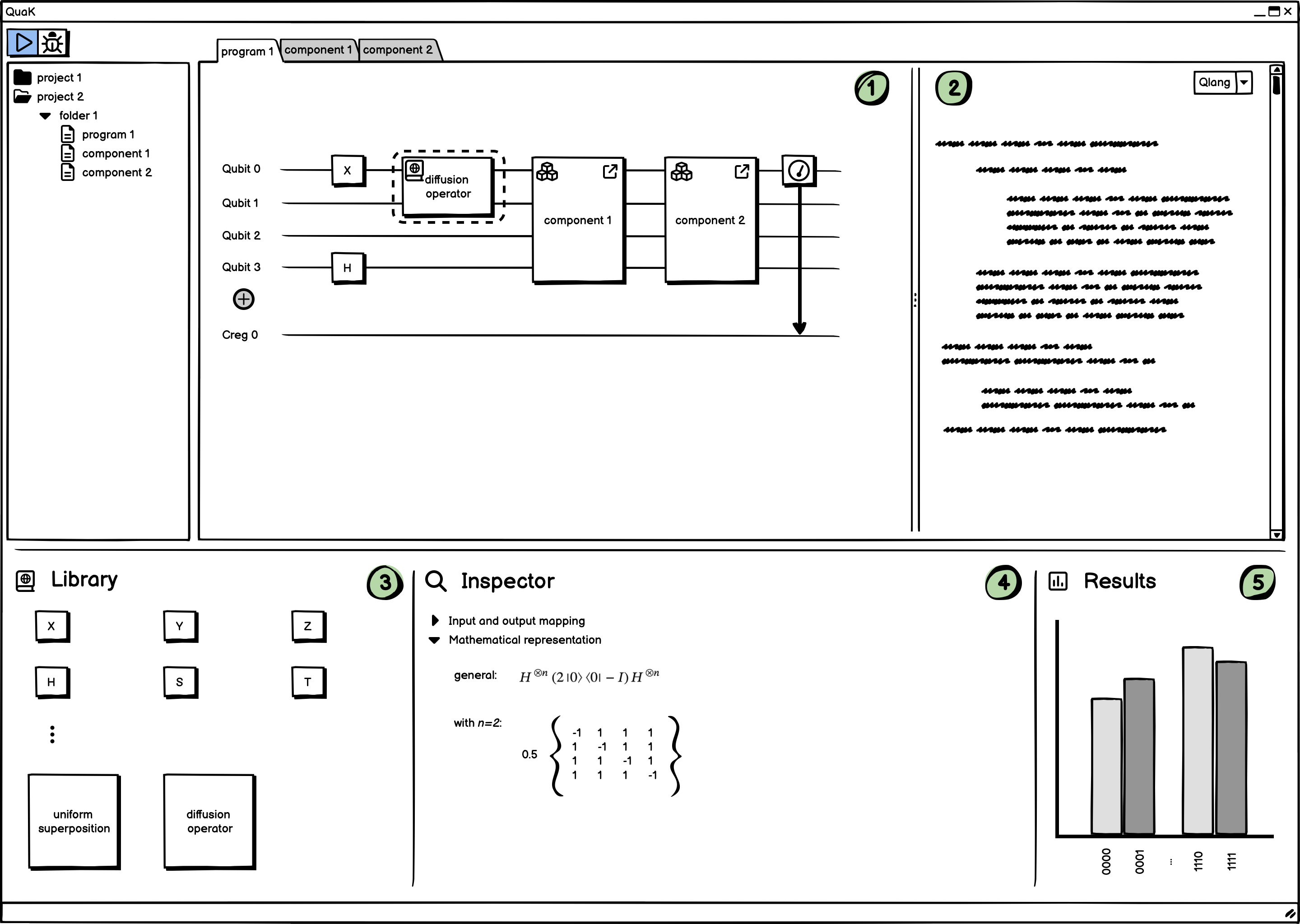}
    \caption{Vision of an IDE for the View-based Quantum Development approach}
    \label{fig:mock}
\end{figure}

We propose a View-based Quantum Development approach based on the V-SUM to comprise different views of quantum computing suitable for stakeholders with varying backgrounds.
Fig.~\ref{fig:mock} shows the approach based on the Quantum Single Underlying Model (Q-SUM), a V-SUM~\cite{meierClassifyingApproachesConstructing2020} tailored for quantum software, as the central artifact.
This model contains cross-disciplinary information such as mathematical descriptions, quantum circuits, and program code.
The Q-SUM allows information to be kept consistent between views, while quantum-specific views can be created dynamically via projection.
Also, only the integration into the Q-SUM is required, for instance, when adding support for a new QPL. 
Quantum Assembly (QASM) can be used as an interface to quantum compilers that handle optimization for the specific quantum hardware.
This approach reuses existing concepts from view- and model-based development and takes advantage of their benefits.


\section{Quantum IDE for the View-based Quantum Development Approach}
A View-based Quantum Development approach requires tool support in order to be applicable in practice.
We propose a Quantum IDE based on the Q-SUM for the collaborative view-based development of quantum software.
Fig.~\ref{fig:mock} depicts how we envision such an IDE.
A rich circuit editor~\circled{1} and QPL editor~\circled{2} should be provided for editing of quantum circuits and quantum program code respectively.
Our IDE addresses open research questions in quantum software identified by~\cite{aliNeedQuantumOrientedParadigm2023}: encapsulation, abstraction, modularity, and quantum platform independence.
Standard gates and more complicated building blocks of quantum software (e.g., a diffusion operator) could be accessed from a library~\circled{3}.
The creation of user-defined blocks and the composition of building blocks should also be possible.
An inspector~\circled{4} should provide a mathematical description and allow configuration of the mapping of connections (e.g., quantum variables as parameters) for a selected building block.
The execution~\circled{5} of the modeled quantum software on quantum hardware and the simulation and inspection of the system state at different points in time should be supported.
For this constraints on the simulation and execution of quantum software must be considered.
The IDE vision is a representative example based on currently employed means of abstraction, but finding appropriate abstractions is an important part of the endeavor.
For example, one might consider a diagrammatic language~\cite{coeckeQuantumPicturalism2010} alternatively to the circuit representation.

\section{Research Challenges}

Realizing this View-based Quantum Development approach with a supporting IDE will require solving the following research challenges:
The conceptual challenge is the design of the Q-SUM that manages the consistency of different views of quantum software based on existing view- and model-based technologies.
To design such a model, it is necessary to identify and analyze the various possible views, and their respective benefits and drawbacks.
The design has to be extensible to integrate existing and future QPLs and support design patterns, best practices,  modularity, and compositionality.
Implementation challenges are the design and implementation of the IDE using a selection of state-of-the-art tooling, and the integration of external quantum simulators and interfaces to quantum hardware.
For the evaluation of the Q-SUM, use case scenarios have to be defined and the capabilities of the approach have to be evaluated against them.
Finally, a user study with domain experts should be conducted to rate the usability of our approach and its supporting IDE.
    

\section*{Acknowledgements}
This work has been supported by the German Ministry for Education and Research in project QuBRA under Grant No.: 13N16303, and in TAQO-PAM under Grant No.: 13NI6092.

\printbibliography

\end{document}